\documentclass[a4paper,10pt,prl,twocolumn]{revtex4-1}
\usepackage{color,amsmath,amssymb,amsthm,times,graphics,graphicx,bm,bbm,dcolumn}

\begin{document}
\title{Prediction of extreme events in the OFC model on a small world network}

\author{Filippo Caruso$^{1,2}$ and Holger Kantz$^3$}
\affiliation{$^1$ Institut f\"{u}r Theoretische Physik, Albert-Einstein-Allee 11, Universit\"{a}t Ulm, D-89069 Ulm, Germany\\
$^2$ QOLS, The Blackett Laboratory, Prince Consort Road, Imperial College, London, SW7 2BW, UK\\
$^3$ Max Planck Institute for the Physics of Complex Systems, N\"othnitzer Str.\ 38, D-01187 Dresden, Germany}

\begin{abstract}
We investigate the predictability of extreme events in a
  dissipative Olami-Feder-Christensen model on a small world topology.
  Due to the mechanism of self-organized criticality, it is
  impossible to predict the magnitude of the next event knowing previous
  ones, if the system has an infinite size.
  However, by exploiting the finite size effects, we
  show that probabilistic predictions of the occurrence of
  extreme events in the next time step are possible in a finite system.
  In particular, the finiteness of the
  system unavoidably leads to repulsive temporal
  correlations of extreme events. The predictability of those is higher for
  larger magnitudes and for larger complex network sizes.
  Finally, we show that our
  prediction analysis is also robust by remarkably reducing the accessible
  number of events used to construct the optimal predictor.
\end{abstract}

\maketitle

\paragraph{Introduction.--}
Self-organized criticality (SOC) is a mechanism by which a large class of
spatially extended dynamical systems, starting in a non-equilibrium
uncorrelated state, can spontaneously organize into a dynamical critical state
with a high degree of correlations \cite{jen_book}. In particular, a partial
synchronization of the elements of the system is usually responsible for
building up long range spatial correlations and thereby
 creating a critical
state, in the thermodynamic limit. The emergence of this SOC complex behaviour
in physical systems is manifested, for instance, by temporal and spatial scale
invariance, i.e. power law or scale free behaviour. SOC has been proposed as a
way to model the widespread occurrence of power laws in nature, i.e. the
abundance of long-range correlations in space and time -- similar to those
observed in critical phase
 transitions \cite{stanley} -- of completely
different events, e.g. luminosity of quasars, chemical reactions, evolution,
sand-pile models, earthquakes, avalanches, forest burns, heart attacks, market
crushes, etc. \cite{BTW,bak_book,jen_book}. Actually, the realistic
applicability of SOC models to describe some of these real events is still
debated -- see, for instance,
Refs. \cite{yang,mega,BTW,bak_book,jen_book,com-corral,Olami,lise,sornette}. An
important consequence of the presence of SOC scenario is that
 in the
critical state the evolution of the system is completely
unpredictable. Indeed, once a SOC system reaches the critical state,
arbitrarily large events can be generated intrinsically by the dynamics
itself. Another important feature of the SOC behaviour is that no fine-tuning
of some external control parameters is required. For a generic initial
condition, after some time interval to let the system `synchronize', it
drives itself into a critical state. However, in the real world the real
systems have a finite size and this unavoidably induces the presence of some
extra correlations between following events. Such correlations will be used in
the following as a useful source of information on the system evolution and
they will allow us to forecast particular events. Actually, we will show that
there exists some nontrivial predictability for large but finite
system sizes and for the
so-called extreme events. The latter can be defined as large deviations from
the `average' behaviour of a complex system and are normally caused by the
presence of intrinsic dynamical fluctuations. These phenomena and their
statistical properties have been intensively studied in literature
\cite{Schellnhuber,Albeverio} but, however, the mechanisms and dynamics
 underlying these
huge deviations are not always fully clarified.
Moreover, the extreme events play an
important role in nature and in our daily life because they are often
associated to destructive events, e.g. hurricanes, strong earthquakes, etc. In
this respect, the predictability of extreme events is urgently desired but
also intensely debated \cite{debates}.
 In this paper, following the
prediction analysis in sand-pile model in Ref. \cite{kantz}, we will
investigate the predictability of such extreme events, by exploiting the time
correlations induced by the finiteness of the system, in a SOC model on a
complex network recently studied in Refs. \cite{Filippo_creta,carusopre}. In
other words, on the basis of past observed events, it is possible to predict
when the next extreme event will happen and the success probability will
increase for `more' extreme events and for larger systems. The latter seems to
be in contradiction to our claim that predictability relies on finite size
effects, but it is true if the magnitude of avalanches is measured relative
to the size of the largest possible avalanche, so that, when comparing
predictability in systems of different size, we also compare events of
different absolute size.

\paragraph{Model.--}
\begin{figure}[t]
\centerline{\includegraphics[width=1\linewidth]{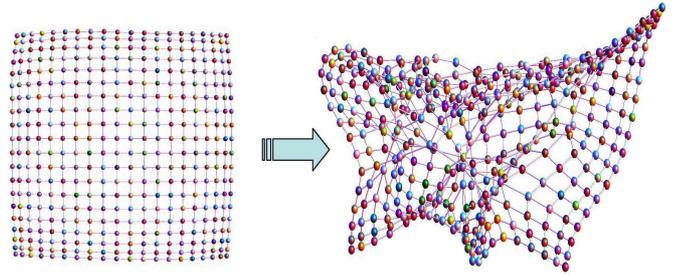}}
\caption{Construction of a small world topology introducing a small fraction
($\sim 2\%$ of the total number of links) of long-range links in a $20$x$20$ square lattice. It corresponds
to a rewiring probability equal to $p_{rew}=0.02$.}\label{fig1}
\end{figure}
Here, we analyze one of the most interesting SOC models,
i.e. the Olami-Feder-Christensen (OFC) model \cite{Olami}. Let us consider a two-dimensional ($L$ x $L$) square lattice of $N=L^2$ sites,
in which each of them carries a real variable  $F_i$ (initially a random value
in $[0,F_{th}]$), representing for instance a seismogenic force, and is
connected by a link to its $4$ nearest neighbours. Then, all these variables
$F_i$ are uniformly and simultaneously increased in order to describe a
loading or stress accumulation (e.g., uniform tectonic loading). At some
point, when one site becomes unstable, i.e. $F_i \geq F_{th}$ with
$F_{th}$ being some threshold value, the driving is stopped (no extra time
step) and a domino effect happens (discharging or stress release), i.e.\ an
earthquake or avalanche starts. In other terms, one has:

\begin{equation}
 \label{av_dyn}
     F_i \geq F_{th}  \Rightarrow \left\{ \begin{array}{l}
      F_i \rightarrow 0 \\
      F_{nn} \rightarrow F_{nn} + \alpha F_i
\end{array} \right.
\end{equation}
with $nn$ being the set of nearest-neighbor sites of $i$. The magnitude of the domino event will be so
defined by the number of these topplings at time step $i$, i.e. $s_i$, while $\alpha$ describes the presence of dissipation in the model.
In the case of a square lattice, the dynamics is conservative for $\alpha=0.25$,
while it is dissipative for $\alpha<0.25$. Here, we will consider a more realistic version of this model, studied recently in Ref. \cite{carusopre,Filippo_creta}, in which a small fraction of long-range links was introduced to obtain a small world topology \cite{watts}.
In particular, the presence of a few long-range edges is enough to synchronize the system and both finite-size
scaling and universal exponents appear \cite{Filippo_creta}. The use of a
small-world topology is justified also by the fact that long-range spatial
correlations have been observed in nature, for instance in earthquake
triggering and interaction, where the static stress may involve relaxation
processes in the asthenosphere with relevant spatial and temporal long-range
effects
\cite{marsan,casarot,cresce,parsons,kagan,turcotte,palatella,abe,corral1,tosi,varotsos}.
The construction of the small world network from a square lattice of size $L$ is explained in more details in Refs. \cite{carusopre,Filippo_creta}. An example of the small world topology construction is qualitatively shown in Fig. (\ref{fig1}). Basically, the links of the lattice are rewired at random with a
probability $p_{rew}$ and small world transition and self-organized criticality (with finite size scaling with
universal critical exponents for the avalanche size probability distribution) are
observed at  $p_{rew}=0.02$, in the dissipative regime, i.e. $\alpha=0.21$ \cite{note}. Notice that, in the
thermodynamical limit, the size probability distribution is a power-law
(without cutoff) and any avalanche size is possible, so both the largest and
the mean avalanche size are infinite. However, in a finite size system,
i.e. for a finite $L$, this probability distribution is a power law with a
cutoff, induced by the finite extension of the lattice,
and we will indicate with
$s_{max}$ the largest avalanche size. This quantity will be numerically
determined in the following. For a generic initial condition, the OFC model on
the small world topology, after a transient (discarded later) to build up
spatial long-range correlations, reaches a critical state and generate a time
series of avalanche size $\{s_i\}$, $i=1,\ldots,n$. In particular, we will
analyze a time series of $n=10^8$ events. Moreover, we define the recurrence
time of avalanches of size larger or equal than $\bar{s}$ as the time interval
($j-i$) in between two consecutive events $s_i$ and $s_j$ such that $s_i \geq
\bar{s}$ and $s_j \geq \bar{s}$, with $j > i$. For infinite lattice size
(thermodynamical limit), the recurrence time distribution is an exponential
for any $\bar{s}$ since successive events are always uncorrelated to each
other. However, as pointed out above, any real system and any numerical
simulation involves a finite size and this thermodynamical limit behaviour
does hold only for small avalanches ($\bar{s} \ll L^2$), which occur in the
bulk of the lattice and are not affected by the presence of
boundaries. Indeed, the recurrence time distribution of large avalanches will
be not exponential but actually one observes the suppression of
short time intervals. The reason of this behaviour is the following. After a
very large avalanche, the system may relax into a sub-critical state and
several time steps are necessary to drive it again into a critical
regime. This effect in turns implies some sort of repulsion
between `extreme' avalanches. These repulsive correlations are quite weak but
useful enough to make predictions on the future occurrence of extreme events.

\begin{figure}[t]
\centerline{\includegraphics[width=1.\linewidth]{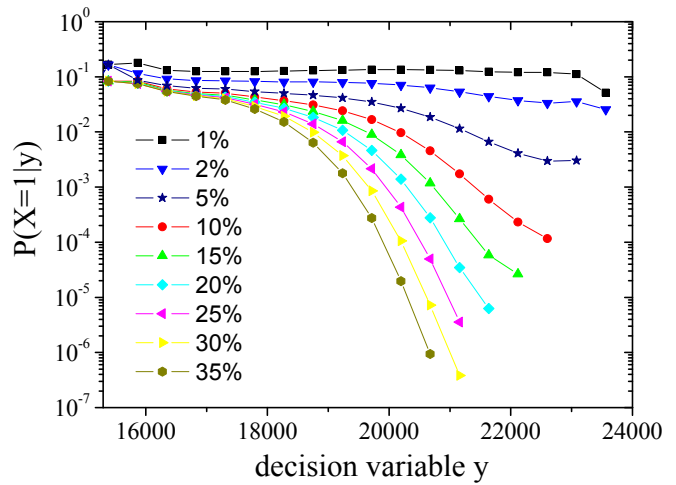}}
\caption{Conditional probability $P(X=1)|y)$ as a function of the decision variable $y$, for
events with size exceeding $1\%$, $2\%$, $5\%$, $10\%$, $15\%$, $20\%$, $25\%$, $30\%$, and
$35\%$ of the maximal avalanche size $s_{max}$, in the case of $L=128$.}\label{fig2}
\end{figure}
\paragraph{Prediction algorithm.--}
Following Ref. \cite{kantz}, we introduce a Boolean series associated to the
events with a size larger than some threshold $\eta$, i.e. $\{X_i\}$ with
$i=1,\ldots, n$ such that $X_i=1$ if $s_i>\eta$ and $X_i=0$ otherwise. The
idea is that, by using information from the past, e.g. from one half of the
time series ($i=1,\dots,n/2$), we want to predict if the next event variable
$X_{j}$ (with $j=n/2+1,\dots,n$) is $1$ or $0$, i.e. whether the next event
will exceed the size $\eta$. In order to do that, we define a decision
variable series $\{y_i\}$ as follows
\begin{equation}
\label{dec-var}
y_i=\sum_{k=1}^i a^k s_{i-k}, \quad 0<a<1,
\end{equation}
with $a$ being a time scale tuning the quality of the prediction. Here, for any threshold $\eta$, as in Ref. \cite{kantz}, $a$ will be chosen of the form $a=\exp\{-1/T(L)\}$ with $T(L=16)=56$, $T(L=32)=112$, $T(L=64)=225$, and $T(L=128)=500$, which are numerically optimized. Notice that in the case, for instance, of $L=128$, $a$ is such that the $1000th$ event before the time $i$, i.e. $k=1000$ in Eq. (\ref{dec-var}), has a weight $1\%$ smaller than the one used for much smaller $k$. In other words, the time scale of the repulsive correlations, discussed above, can be at most considered of the order of thousands events. Besides, let us stress that $y_i$ does depend only on the past events and so from the knowledge of the past time series one wants to forecast the future events.
As discussed in Ref. \cite{kantz}, the possibility of making predictions is subordinated to the fact that the conditional probability
$P(X=1|y)$, i.e. the probability for an avalanche with size larger than $\eta$ to occur given a specific value of the decision variable $y$, must show at least a maximum as a function of $y$. This necessary condition for predictability is investigated in our case for several values of $\eta$ in terms of $s_{max}$, as shown in Fig. (\ref{fig2}). We find that for small sizes, i.e. $\eta$ equal to a few $\%$ of $s_{max}$, the conditional probability $P(X=1|y)$ has almost a flat behaviour, implying that the probability for an event to happen is completely independent on the past, i.e. no predictability is possible. In this case, the finite-size effects are irrelevant and the behaviour of small avalanches well approximates the temporal features of the thermodynamical limit. However, for large avalanche sizes $P(X=1|y)$ shows a maximum at small values of $y$, i.e. it is more likely to have a large event when only small avalanches happened in the recent past. It is another way to show that extreme events repel each other and between two consecutive ones the system is in a sub-critical state and generates mainly small avalanches. Let us point out also that the maximum of $P(X=1|y)$ is more pronounced for larger $\eta$, i.e. larger events are more predictable.
Now, we will describe in a more quantitative way how to make predictions once one knows the decision variable $y$. In particular, one needs to define
a probabilistic predictor on $\{y_i\}$, defined as a map $y_i \mapsto \hat{p}_i$, with $\hat{p}_i\in [0,1]$. It can be proved that the optimal predictor is simply characterized by the conditional probability, i.e. $\hat{p}_i=P(X=1|y_i)$ \cite{Neyman,Hallerberg,kantz}. In general, this predictor is probabilistic, since it is defined by $P(X=1|y_i)$, but it can converted into a deterministic one by simply introducing a threshold, as follows
\begin{eqnarray}
\label{det-pred}
X^{pred}_i & = & \left\{ \begin{array}{ll}
1\quad :& \mbox{if} \quad P(X=1|y_i)> r_{alarm}, \\
0\quad :& \mbox{otherwise,}
\end{array} \right.
\end{eqnarray}
with $r_{alarm}$ being a control parameter determining the total alarm rate, while $X^{pred}_i$ is the predicted value of the Boolean variable $X_i$.
By increasing the value of $r_{alarm}$, the total alarm rate decreases,
i.e. one predicts one event only when the occurrence probability is really
high. Of course, \textit{a posteriori} we may check whether our prediction is
right or false by just comparing the variables $X_i$ and $X^{pred}_i$. On one
hand, when $X_i=X^{pred}_i=1$, the prediction is correct and this event is
counted in the so-called hit rate. On the other hand, when $X_i \neq
X^{pred}_i=1$, the prediction is wrong and it does correspond to a false
alarm. Notice also that, if $r_{alarm}=0$, the condition in (\ref{det-pred})
is always satisfied and both rates tend asymptotically to $1$. Instead, for
$r_{alarm}$ close to the maximum of $P(X=1|y_i)$, that inequality does never
hold and both rates are zero, i.e. one does not make any
prediction.

Specifically, we will use the first half of our avalanche time
series of $10^8$ events as a training set to construct the predictor,
which means that we estimate $P(X=1|y_i)$ on these data.
We then feed this $P(X=1|y_i)$ with $y_i$ values from the second half
and compare the outcome to the corresponding boolean values $X_i$.
\paragraph{Prediction quality and results.--}
\begin{figure}[t]
\centerline{\includegraphics[width=.97\linewidth]{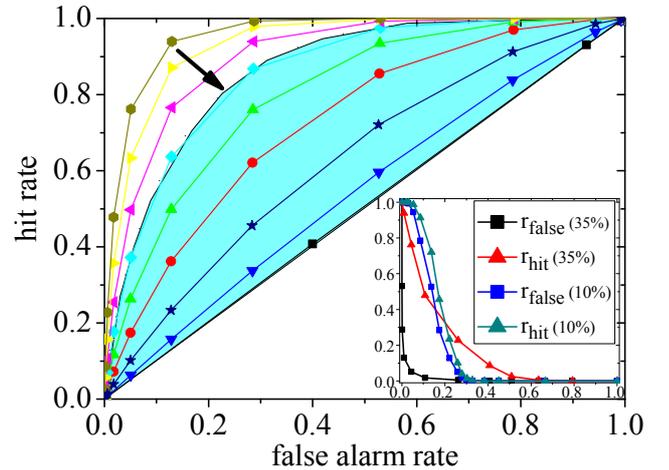}}
\caption{ROC plots for the prediction of extreme events, for
events with size exceeding $1\%$, $2\%$, $5\%$, $10\%$, $15\%$, $20\%$, $25\%$, $30\%$, and
$35\%$ of the maximal avalanche size $s_{max}$ (from bottom to top), in the case of $L=128$.
These curves are squeezed to the colored (blue) area in the case of $L=64$, i.e. the predictability is better for larger system sizes. This behaviour does hold also for other smaller lattice sizes, that we have considered, e.g. $L=16$ and $L=32$.
Inset: Hit rate and false alarm rate versus the threshold $r_{alarm}$ (normalized to $\max[P(X=1|y)]$), for $\eta=10\%, 35\% \ s_{max}$.}\label{fig3}
\end{figure}
Here, we analyze in more detail the quality of the $y$-based predictions in
the case of the OFC model on a small world network, described above.
 When we
use the prediction algorithm, there are two possible errors ($X_i \neq
X^{pred}_i$) : 1) missing an event, i.e. $X_i=1$ and $X^{pred}_i=0$, ii) false
alarm, i.e. $X_i=0$ and $X^{pred}_i=1$. A possible method to characterize the
prediction quality is the Receiver Operating Characteristics (ROC)
\cite{Egan}, which was recently used in Ref. \cite{kantz} for a prediction
analysis in a sand-pile model. It consists of comparing (in ROC plot) the hit
rate $r_{hit}$ and the false alarm rate $r_{false}$, as a function of the
threshold $r_{alarm}$. The benchmark is the case $r_{hit}=r_{false}$,
i.e. the diagonal in the ROC plot. This is the outcome if predictions
 $X^{pred}_i=1$ are made at random times, independent of the values of $y_i$.
When $r_{hit} > r_{false}$, the predictor is
useful and the distance of the ROC curve from the benchmark diagonal is a good
indicator of the quality of the prediction. The ROC curves for the model above
in the case of $L=128$ are shown in Fig. (\ref{fig3}) for different thresholds
$\eta$ in terms of the maximum avalanche size $s_{max}$. Note that, increasing
the value of $r_{alarm}$, one can get a very large hit rate with a small false
alarm rate (i.e., very good prediction). The extreme points of each ROC curve
corresponds to the trivial situations: a) $r_{alarm}=0$, with
$r_{hit}=r_{false}=1$, b) $r_{alarm}=\max P(X=1|y_i)$ with
$r_{hit}=r_{false}=0$. Let us stress that the predictability is better for
larger threshold avalanche sizes. In the inset of Fig. \ref{fig3}, we show the
behaviour of the hit rate and the false alarm rate as a function of
$r_{alarm}$ normalized to the maximum value of the conditional
probability $P(X=1|y)$. When lowering the threshold $r_{alarm}$, the
sensitivity increases, and for large avalanches ($\eta$ = 35\%),
the hit rate starts to rise much earlier than the false alarm rate, indicating
the predictive skill.

\begin{figure}[t]
\centerline{\includegraphics[width=.95\linewidth]{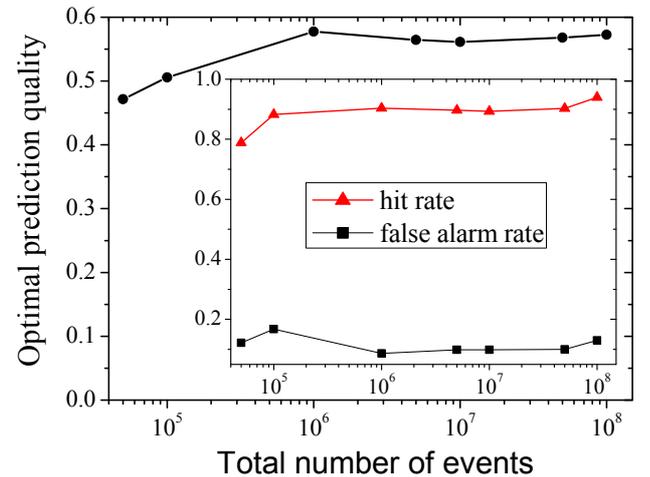}}
\caption{Optimal prediction quality, measured as the largest distance of the ROC curve from the benchmark ($r_{hit}=r_{false}$), versus $n$, in the case of $L=128$ and $\eta=35\% \ s_{max}$. Inset: Optimal hit and false alarm rates versus $n$. The predictability is not affected by remarkably reducing the number of events used to define the decision variable.}\label{fig4}
\end{figure}

We have applied this analysis also to
smaller system sizes, e.g. $L=16,\; 32,\; 64$. There are two essential
results: For given system size, we always detect some nontrivial predictability
of large events, and the predictability is the better the larger the events we
intend to predict. This is an evident consequence of the fact that these
are more strongly affected by the finiteness of the network. When comparing
different system sizes and predicting events of identical magnitude,
the smaller system is better predictable. In this respect, the predictability
asymptotically vanishes for infinite system size, as expected.
If, however, we measure event
magnitudes as percent of $s_{max}$, then larger systems are better
predictable. For instance, for
$L=64$, all ROC curves are squeezed to the blue area in Fig. (\ref{fig3}) and
the prediction quality is lower that in the case of $L=128$.
Finally, we consider how the
predictability quality changes by varying the total number of events in the
time series, i.e. $n$. Indeed, a so large number of events, like $10^8$, could
be not accessible for real phenomena. However, in Fig. \ref{fig4}, we show the
optimal prediction quality, measured as the largest distance of the ROC curve
from the benchmark ($r_{hit}=r_{false}$), and the corresponding hit rate and
false alarm rate as a function of the total number of events $n$. We find that
the predictability is similar even constructing the decision variable on a
number of events which is three orders of magnitude smaller than the one used
above. In other words, this prediction analysis seems to be also quite robust
with respect to the size of the accessible sample.
\paragraph{Conclusions and Outlook.--}
We have investigated the time series of extreme events generated by a dissipative Olami-Feder-Christensen model on a small world
network. The small world property is here essential to
create vary large events - without small world, the system is not truly
critical \cite{carusopre}.
The presence of finite-size effects induces repulsive time correlations between consecutive extreme events and this can be used to make predictions. In particular, we have considered a decision variable which keeps record only of the recently past events and, by using the conditional probability as optimal predictor, we have shown that the predictability quality is really good for large avalanche sizes and for larger networks. In this respect, we have applied a ROC analysis and found that the ratio hit rate/false alarm rate can be remarkably high when considering `more extreme' events. Therefore, these results show that, although the SOC models can be applied to describe very well real events (implying also that they are uncorrelated in time), however, the fact that real physical systems and practical models have a finite size can be exploited to extract information from the extra temporal correlations, induced by the finite-size effects, and forecast the occurrence of extreme events. Interestingly enough, the quality of the predictability is higher for larger systems and for more `catastrophic' events, whose predictability is, of course, even more urgently desired. Finally, although maybe most real phenomena may not share the properties of the SOC model analyzed above, our results suggest to exploit the presence of, though weak, temporal correlations (not necessarily coming from finite-size effects, but also from other sources), to try to forecast extreme events.

\acknowledgments
F.C. thanks A. Pluchino and A. Rapisarda for discussions.
This work was supported also by a Marie Curie Intra European Fellowship within
the 7th European Community Framework Programme.


\begin{thebibliography}{0}
\bibitem{note}
Note that the presence of SOC behaviour in the dissipative OFC model on small world network was verified also for other values of $\alpha$, i.e.
$0.15 \leq \alpha \leq 0.21$, in Ref. \cite{carusopre}.
\end{thebibliography}
\end{document}